\documentclass[aps,pre,twocolumn,showpacs,groupeaddress]{revtex4}
\usepackage{braket}
\usepackage{comment} 
\usepackage{graphicx}
\usepackage{dcolumn}
\usepackage{bm}
\begin{document}
\title{Ground-state properties of the one-dimensional transverse Ising model in a longitudinal magnetic field}
\author{O. F. \surname{de Alcantara Bonfim}}
\email{bonfim@up.edu}
\affiliation{Department of Physics, University of Portland, Portland, Oregon 97203, USA}
\author{B. \surname{Boechat}}
\email{bmbp@if.uff.br}
\affiliation{Departamento de F\'{\i}sica, Universidade Federal Fluminense\\
Av. Litor\^anea s/n,  Niter\'oi, 24210-340, RJ, Brazil}
\author{J. \surname{Florencio}}
\email{jfj@if.uff.br}
\affiliation{Departamento de F\'{\i}sica, Universidade Federal Fluminense\\
Av. Litor\^anea s/n,  Niter\'oi, 24210-340, RJ, Brazil}

\date{\today}

\begin{abstract}

\vskip 12pt

The critical properties of the one-dimensional spin-$1/2$ transverse Ising model in the presence 
of a longitudinal magnetic field were studied by the quantum fidelity method. We used exact diagonalization 
to obtain the ground-state energies and corresponding eigenvectors for lattice sizes up to 24 spins. 
The maximum of the fidelity susceptibility was used to locate the various phase boundaries present in 
the system. The type of dominant spin ordering for each phase was identified by examining the 
 corresponding ground-state eigenvector. For a given antiferromagnetic nearest-neighbor interaction $J_2$, 
we calculated the  fidelity susceptibility as a function of the transverse field $B_x$ and the  longitudinal 
field $B_z$. The phase diagram in the  ($B_x,B_z$)-plane shows three phases. These findings are 
in contrast with the published literature that claims that the system has only two phases.
For $B_x < 1$, we observed an antiferromagnetic phase for small values of $B_z $ and a paramagnetic 
phase for large values of $B_z$. For $B_x > 1$ and low $B_z$, we found a disordered phase that undergoes 
a second-order phase transition to a paramagnetic phase for large values of $B_z$.
\end{abstract}


\maketitle

\section {\label{sec:I} Introduction}

There is an ongoing interest in the zero-temperature properties of quantum spin systems~\cite{Sac11,Yus18}.
 In particular on the nature of phase transitions that
occur due to the presence of pure quantum fluctuations since thermal fluctuations 
are absent. These transitions are triggered when a  Hamiltonian
parameter crosses a given value, upon which there occurs a change 
in the spin arrangement in the underlying lattice.  Such transitions are regarded 
as second-order when the changes of the ground-state properties are continuous.
On the other hand, if the changes are discontinuous 
the system undergoes a first-order transition.  All of these 
can only occur at the thermodynamic limit, when the size of the
system is infinite.    

The one-dimensional (1D) transverse Ising model in a longitudinal 
field is a relatively simple model that displays both continuous 
and discontinuous transitions. Thus it has drawn a considerable amount 
of interest in the literature.    
Several approaches have been used to study that model, namely, 
entanglement measures~\cite{Yus18},  simulations with
ultra-cold atoms in optical lattices~\cite{Sim11,Lew07,Blo08,Gre02},
density matrix renormalization group (DMRG)~\cite{Ovc03,Cam14,Rob17,Pel18},
quantum Monte Carlo~\cite{Nov86}, neural networks~\cite{Czi18}, exact
diagonalization~\cite{Sen00,Ban11}, and finite size scaling~\cite{Cam14,Ros18}.
In  numerical calculations, finite-size scaling is often employed 
to infer the location of the transitions in the thermodynamic limit.

In the present work, we use the fidelity method to find the
zero-temperature phase diagram of the 1D spin-$1/2$ transverse Ising
model in a longitudinal field.
That method  is well suited to the identification of phase changes, 
as it relies upon the detailed properties of the ground-state 
eigenvectors~\cite{Ben92,Aba08,Coz07,Bon17}.
It is very sensitive to changes in the quantum state
of the system, and provides precise information about the location
of the phase transitions as a given Hamiltonian parameter
is varied.  The nature of the transition can also be
determined  by the method.
It has been used to detect and characterize a variety of phase transitions 
without requiring prior knowledge of the local order parameter of the system. 
This point of view also leads to new ways of looking at phase transitions and reveals 
the origin of their universalities.

Due to its simplicity and ability to locate phase transitions, quantum fidelity has been used 
in quantum information theory~\cite{Ben92} and for the identification of topological phases 
in condensed matter physics~\cite{Aba08,Coz07}. A unified approach connecting Berry 
phases and quantum fidelity has been established~\cite{Ven07}. Monte Carlo schemes were
introduced to compute the fidelity and its susceptibility for large interacting many-body 
systems in arbitrary dimensions~\cite{Sch09}. An analysis of the transverse Ising model
in the thermodynamic limit shows the universal properties of the fidelity near a critical 
point~\cite{Ram11}. 

The fidelity method has also been used to identify the universality class of the quantum transitions 
in the 1D  asymmetric Hubbard model~\cite{Gu08}.
Scaling relations for the fidelity susceptibility in the quantum critical regime have 
been derived~\cite{Alb10}. The scaling behavior of the fidelity susceptibility in the vicinity
of a quantum multicritical point has been also studied~\cite{Muk11}.
The quantum properties of the two-dimensional version of the present model has been 
investigated by exact diagonalization using both longitudinal and transverse fidelity 
susceptibilities~\cite{Nis13}. An exact expression for the fidelity susceptibility for the Ising 
model in a transverse field has been derived~\cite{Dam13}. Quantum fidelity has been used
to identify ground-state degeneracy of quantum spin systems \cite{Su13}.
The behavior of the ground-state fidelity susceptibility in 1D quantum systems displaying a
Berezinskii-Kosterlitz-Thouless type transition has been also investigated~\cite{Sun15}.
A method to calculate the fidelity susceptibility of correlated bosons, fermions and quantum
spins systems at both zero and non-zero temperatures has been proposed using a variety 
of quantum Monte Carlo methods~\cite{Wan15}. An extension and generalization of the 
application of the fidelity susceptibility to strongly correlated lattices systems has been 
put forward~\cite{Hua16}. Closed-form expressions for the fidelity susceptibility of the 
anisotropic $XY$-model in a transverse field has been recently found~\cite{Luo18}.
 Dynamical phase transitions at finite temperatures have recently been studied in 
topological systems by means of fidelity susceptibility~\cite{Mer18}. 
A comprehensive review of  the fidelity approach to quantum phase 
transitions is given in Ref.~\cite{Gu10}. 

Recently, the fidelity method was  used to study the transverse
Ising model with next-to-nearest neighbor interactions~\cite{Bon17},
where it uncovered other quantum phases whose existence
had been overlooked by other approaches.  Thus, 
the fidelity method helped to uncover  a much richer phase
diagram for that model.  

The phase diagrams found in the literature for the 1D spin-$1/2$ transverse 
Ising model in a longitudinal field show an 
antiferromagnetic phase at low fields and a paramagnetic 
phase at high fields \cite{Yus18,Sim11,Ovc03}.  
There appears a transition line for
a continuous transition belonging to the same
universality class of the 2D Ising model~\cite{Sen00}.
When the transverse field vanishes, 
the model shows a multicritical point 
where a first-order transition occurs.
The phase diagrams in the literature
show a single transition between antiferromagnetic and paramagnetic
phases.  
As we shall see below, our fidelity approach uncovers
 an additional phase boundary line between the paramagnetic 
 phase and a disordered phase, in addition to
reproducing the boundary line found in the literature.

This paper is organized as follows: In Sec.~II  we present the
model, while in Sec.~III we discuss the fidelity susceptibility
method.  In Sec.~IV we present our results and finally, we summarize
our results in Sec.~V.

\section{\label{sec:II} The Model} 

The 1D transverse Ising model in the presence of a longitudinal field is written as
\begin{equation}
{\cal H} = J_2\sum_{i} \sigma^z_{i}\sigma^z_{i+1}  - B_x \sum_{i} \sigma^x_{i}
 -  B_{z} \sum_{i} \sigma^z_{i}.
\label{eq:Hamiltonian}
\end{equation}
The chain consists of $L$ spin-half  interacting spins, written in terms of Pauli operators,
 where $\sigma^{\alpha}_i$ $(\alpha = x,y,z)$ is the $\alpha$-component located at site $i$.  
 We consider a chain with periodic boundary conditions. 
The  nearest-neighbor Ising coupling is antiferromagnetic, that is $J_2 > 0$, while the applied 
longitudinal field $B_{z} > 0$  tends to align the spins ferromagnetically. 
For $B_{z} > 0$ the model is gapped with a non-degenerate ground-state.
Finally, quantum fluctuations are induced by a transverse magnetic field $B_x$. 
In what follows, we take $J_2 = 1$ as the energy unit.

When $B_x = 0$, the Hamiltonian reduces to the Ising model in a longitudinal magnetic field.
In that case the model shows a first-order phase transition at $B_z = 2.0$. 
For $ B_x \ge 0$, the ground state of the system in the low-field regime ($B_z < 2.0$) is 
antiferromagnetic, whereas for high fields ($B_z > 2.0$) it is paramagnetic separated by 
a second-order transition except at  the multicritical point $(B_x, B_z) = (0.0, 2.0)$ 
where the quantum fluctuations are suppressed and a classical first order phase transition 
occurs~\cite{Sim11}.

On the on the hand, for $B_z = 0$, the Hamiltonian becomes  
the transverse Ising model, whose ground-state properties were exactly obtained by 
Pfeuty in 1970~\cite{Pfe70}. 
He found that quantum fluctuations induced by the transverse field drive 
the system through a second-order phase transition at $B_x = 1.0$.
At  low-fields the phase is antiferromagnetic, whereas for high fields it is disordered.

\section{\label{sec:III} The Fidelity Approach}

Consider an Hamiltonian that depends on an arbitrary parameter $\lambda$, which
drives the system through a phase transition when $\lambda= \lambda_c$.
We define the quantum fidelity of a ground-state as the magnitude of the overlap
between two neighboring ground-states, namely,
\begin{equation}
           F(\lambda,\delta) = |\Braket{\psi(\lambda)|\psi(\lambda + \delta)}|,
\label{eq:fidelity}
\end{equation}
where $\Ket{\psi}$ is the normalized non-degenerate ground-state eigenvector 
evaluated near a given value of $\lambda$ by an arbitrary small shift $\delta$. 

Quantum fidelity also depends on the system size.  
As the system approaches a quantum transition, the fidelity
behavior changes dramatically.  It drops from a level close to unity on either side of the 
transition point, to a minimum value at the transition point. This is caused by the distinct 
nature of the ground-state at each side of that transition point.

Instead of working with quantum fidelity as defined above, it is preferable to work with the 
fidelity susceptibility, which is obtained by expanding the fidelity as a Taylor's series for 
very small $\delta$ about $\lambda$. 
Assuming that the ground state is normalized, the fidelity susceptibility can be written as
\begin{equation}
         \chi(\lambda) = 2(1 - F(\lambda,\delta))/\delta^2 +  {\cal{O}}(\delta^2).
\label{eq:susceptibility}
\end{equation}

In the present work, the ground-state energy and eigenvector for a given $\lambda$  are found 
by using both Lanczos and conjugate gradient methods. The latter has been used in 
Hamiltonian models in statistical physics and transfer-matrix techniques ~\cite{Nig90,Nig93}. 
For a given accuracy, both methods give the same results for the ground-state eigenvectors
and energies. 

Since the Hamiltonian (\ref{eq:Hamiltonian})  depends on two independent parameters, 
$B_x$ and $B_z$, we must investigate each of their associated susceptibilities. 
To differentiate between them, we use the notation $\chi_{\gamma}(\lambda)$, 
where $\lambda$ is chosen as one of the fields, and $\gamma$ is the other field, 
which is kept fixed during the calculations. In our numerical calculations, the boundary lines 
are found by using Eq.~(\ref{eq:susceptibility}) with $\delta = 0.001$ with a range of accuracy 
between $10^{-12}$ and $10^{-14}$ for the ground-state energy, depending 
on the chain size. For each $\lambda$, the location of the phase boundary is 
determined by the maximum of the fidelity susceptibility.

We write the Hamiltonian using
the standard basis consisting of a tensor product of $L$ eigenstates of the $z$-component of the 
Pauli operator $\sigma^{z}_i$ located at site $i$, namely   $|n> = \prod_i^L\, |s>_i$.
On each site $i$ we have $s= 0, 1$, where  $|s=1>_i$ denotes 
the eigenvector of $\sigma^{z}_i$ for an up-spin, and $|s=0>_i$ is the corresponding eigenvector 
for a down-spin. 
The index $n$ labels the basis states and has the values 
$n= 0,1,...,N-1$, with $N= 2^L$ which denotes the size of the Hilbert space.

By writing the  basis index $n$ in binary notation, each of the $L$ binary digits will represent 
the z-component of the spin at a given site $i$ of a lattice with $L$ spins. 
An arbitrary eigenstate of the Hamiltonian can therefore be written as:
 \begin{equation}
|\phi_{\alpha}> = \sum_{n=0}^{N-1}a_{\alpha}(n)|n>,
\label{eq:phi}
\end{equation}
where the energy levels are labeled by $\alpha = 0,1, ...,N-1$.  In particular, $\alpha = 0$ 
is assigned to the ground-state.

 Because of the symmetry of the Hamiltonian (\ref{eq:Hamiltonian}), the coefficients  
 $a_{\alpha}(n)$ are real. The full wave vector can be visualized by plotting the amplitudes 
 $a_{\alpha}(n)$ for any lattice size $L$, as a function of the state index $n$ in a single 
graph~\cite{Bon06,Boe14, Bon14}.

\begin{figure}
\includegraphics[width=8.0cm, height= 6.0cm, angle=0]{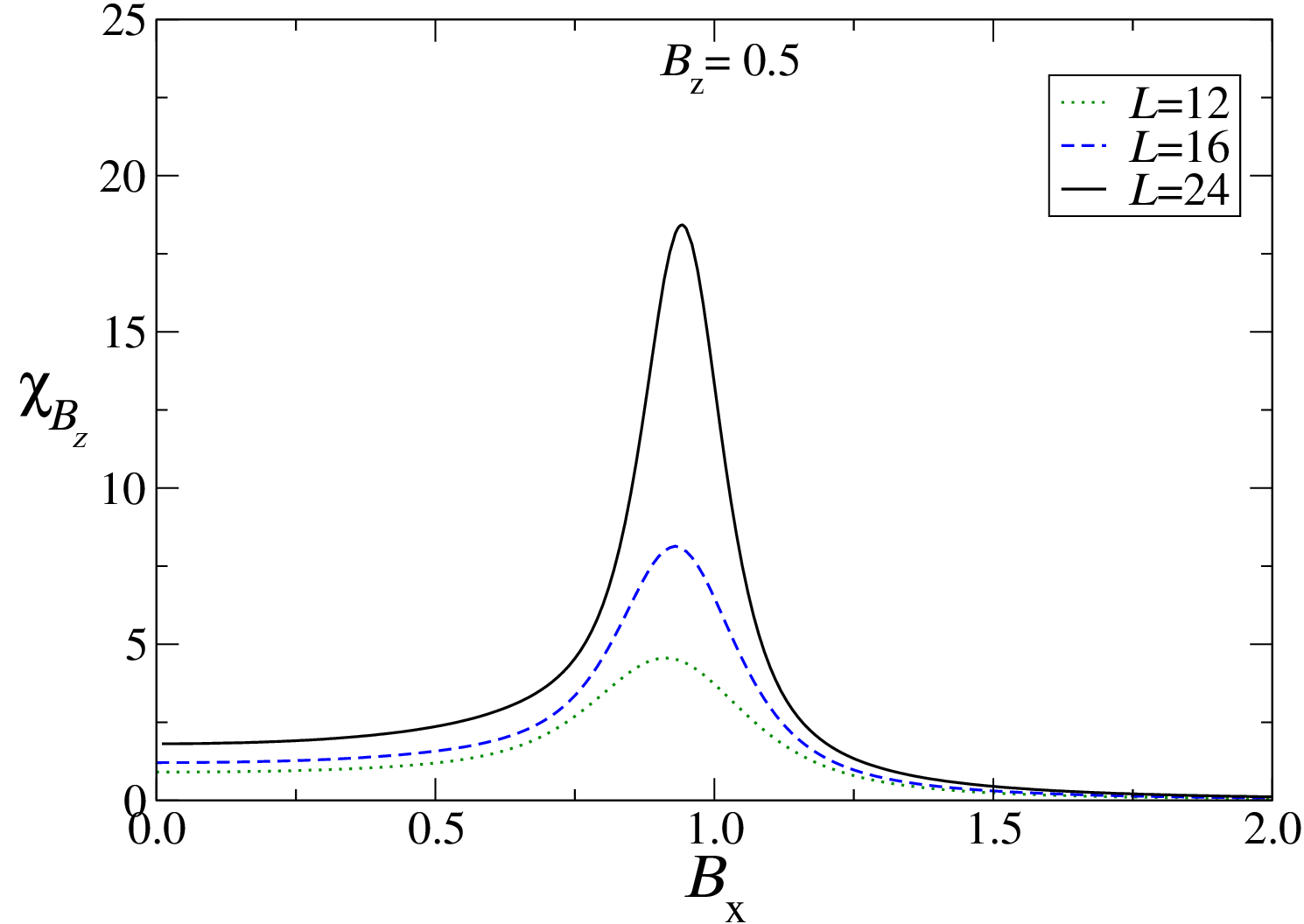}
\setlength{\abovecaptionskip}{0pt}
\setlength{\belowcaptionskip}{40pt}
\caption{(color online) Fidelity susceptibility as a function of the transverse field $B_x$
for a fixed longitudinal field $B_z = 0.5$, and lattice sizes  
$L = 12$, $16$, and $24$. 
The maximum of the susceptibility specifies the location of the transition point. 
We set $J_2= 1$ as the unity of energy for this and the subsequent figures.
\label{fig:figure1}
}
\end{figure}

\begin{figure}
\includegraphics[width=8.0cm, height= 6.0cm, angle=0]{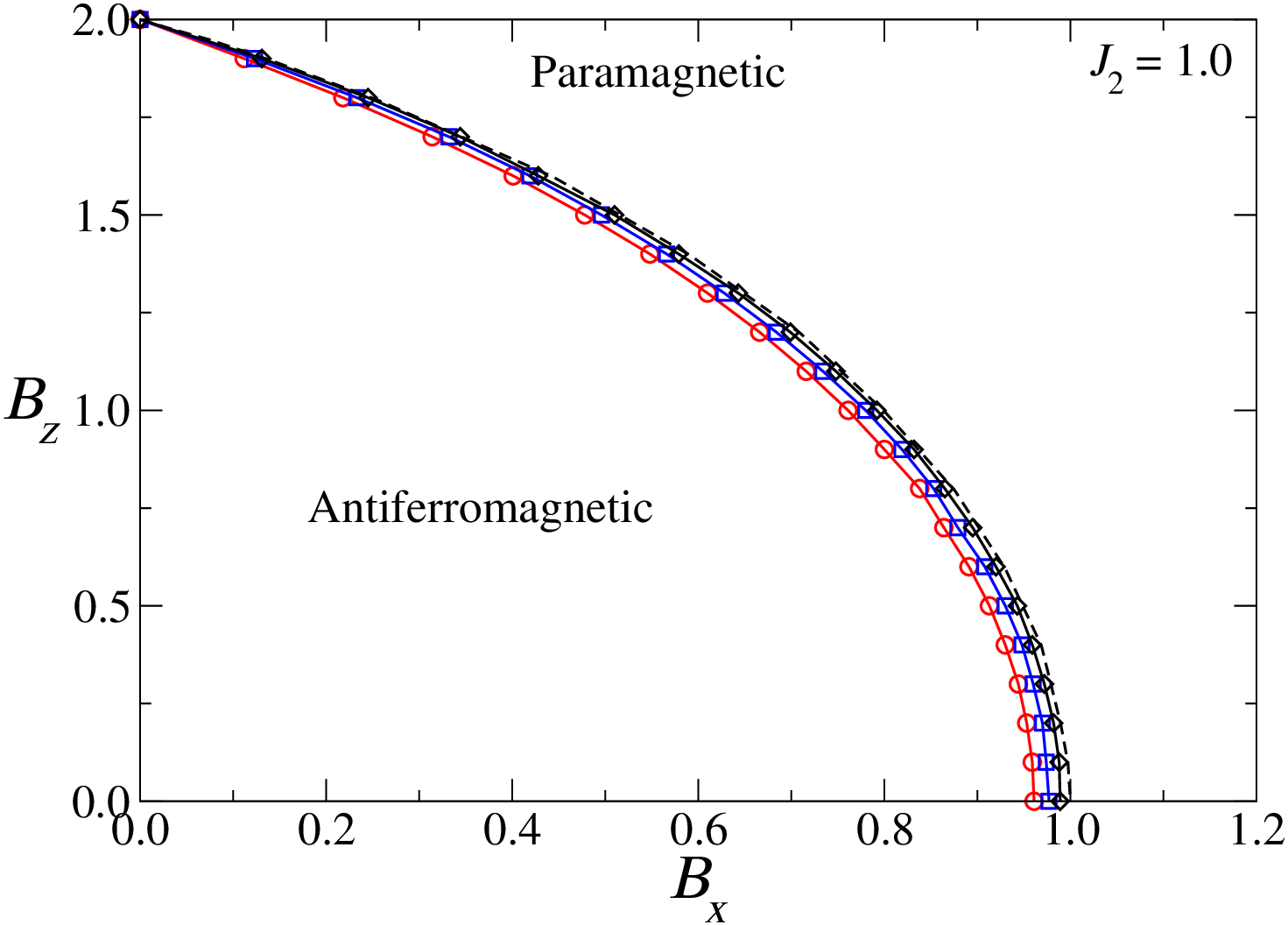}
\setlength{\abovecaptionskip}{0pt}
\setlength{\belowcaptionskip}{40pt}
\caption{(color online) Phase diagram in the $(B_x, B_z)$-plane for chains sizes $L = 12$ (circles), 
$16$ (squares) and $24$ (diamonds), obtained from the maximum of $\chi_{B_z}$.
The model shows two phase regions, antiferromagnetic  and  paramagnetic.
The transition points  $(B_x,B_z)= (0, 2)$ and $(B_x,B_z)= (1, 0)$  are exact results.
The dashed line is the critical line from DMRG results (Ref.~\cite{Ovc03}).
\label{fig:figure2}
}
\end{figure}

\begin{figure}
\includegraphics[width=8.0cm, height= 6.0cm, angle=0]{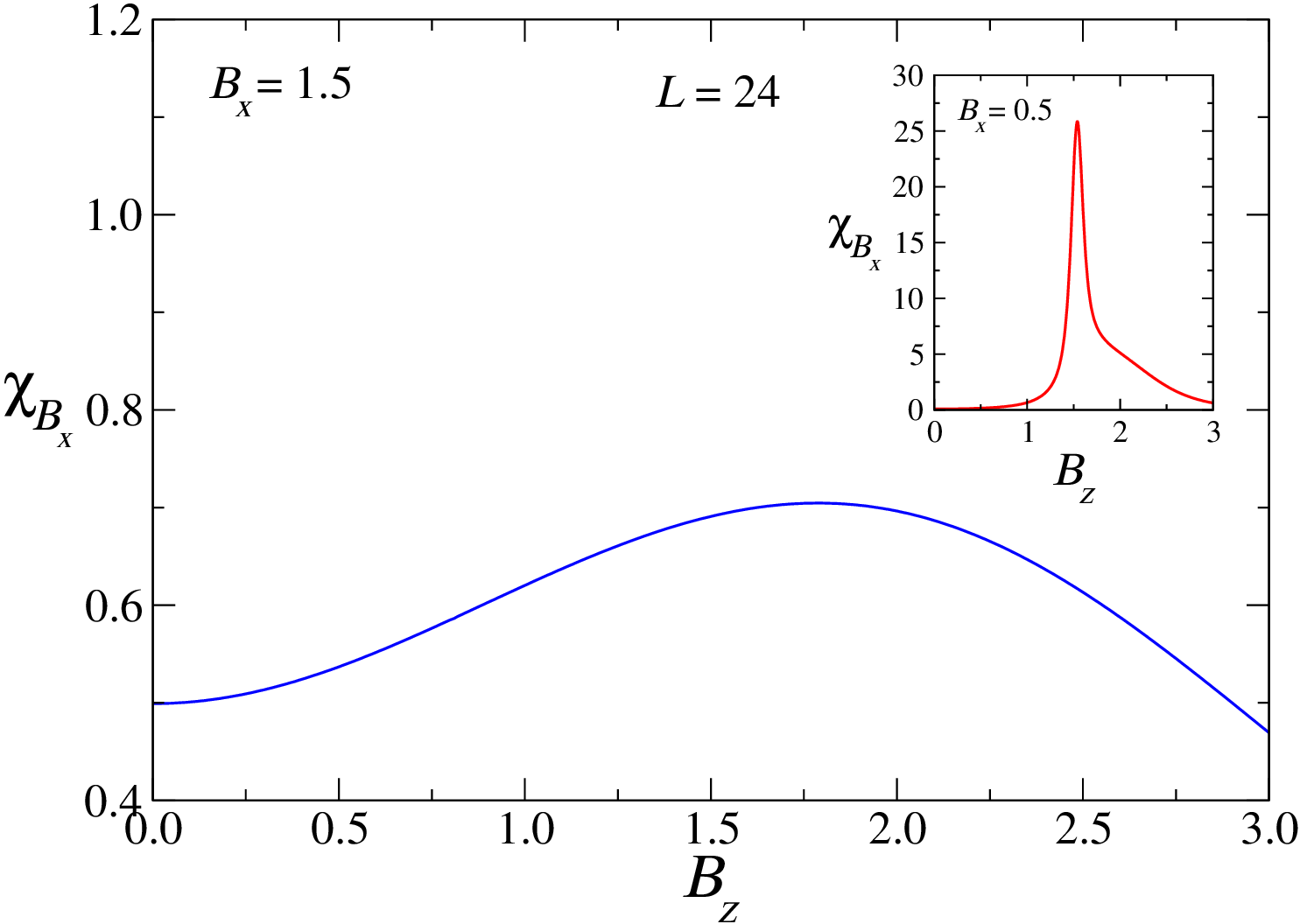}
\setlength{\abovecaptionskip}{0pt}
\setlength{\belowcaptionskip}{40pt}
\caption{(color online) Fidelity susceptibility {\em vs} longitudinal field $B_z$
for a chain of size $L = 24$, and fixed transverse fields $B_x$ = 1.5 and 0.5 (inset). 
The maximum of the susceptibility locates the transition point. 
\label{fig:figure3}
}
\end{figure}

\begin{figure}
\includegraphics[width=8.0cm, height= 6.0cm, angle=0]{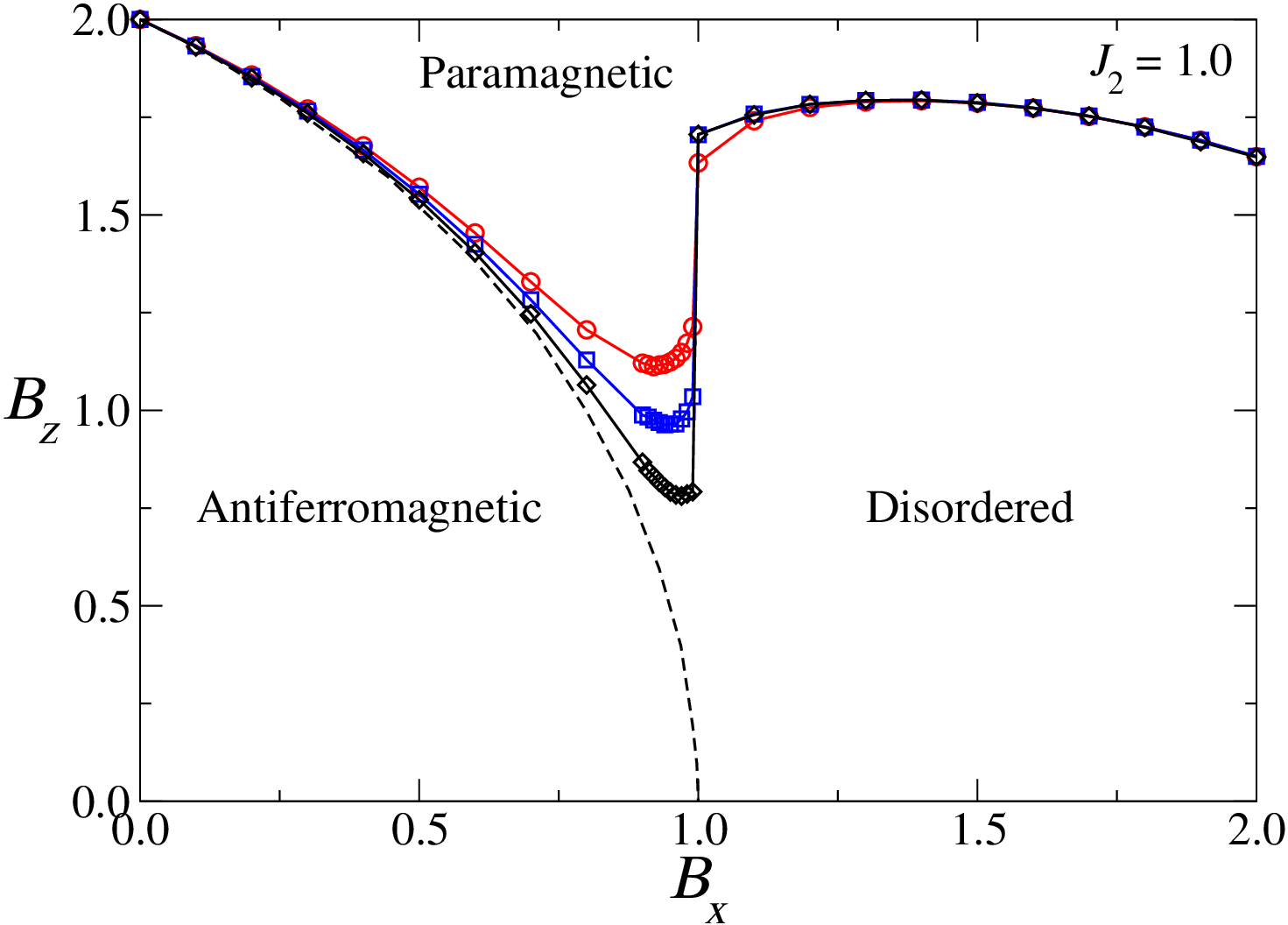}
\setlength{\abovecaptionskip}{0pt}
\setlength{\belowcaptionskip}{40pt}
\caption{(color online) Phase diagram in the $(B_x, B_z)$-plane for chains with $L = 12$ (circles), 
$16$ (squares), and $24$ (diamonds), obtained from the maxima of $\chi_{B_x}$.
The model shows three phase regions, antiferromagnetic, paramagnetic, and disordered.
The transition points  at $(B_x,B_z) = (0, 2)$ and $(B_x,B_z) = (1, 0)$   are known exact results.
The dashed line is the critical line from DMRG (Ref.~\cite{Ovc03}). 
\label{fig:figure4}
}
\end{figure}

\begin{figure}
\includegraphics[width=8.0cm, height= 6.0cm, angle=0]{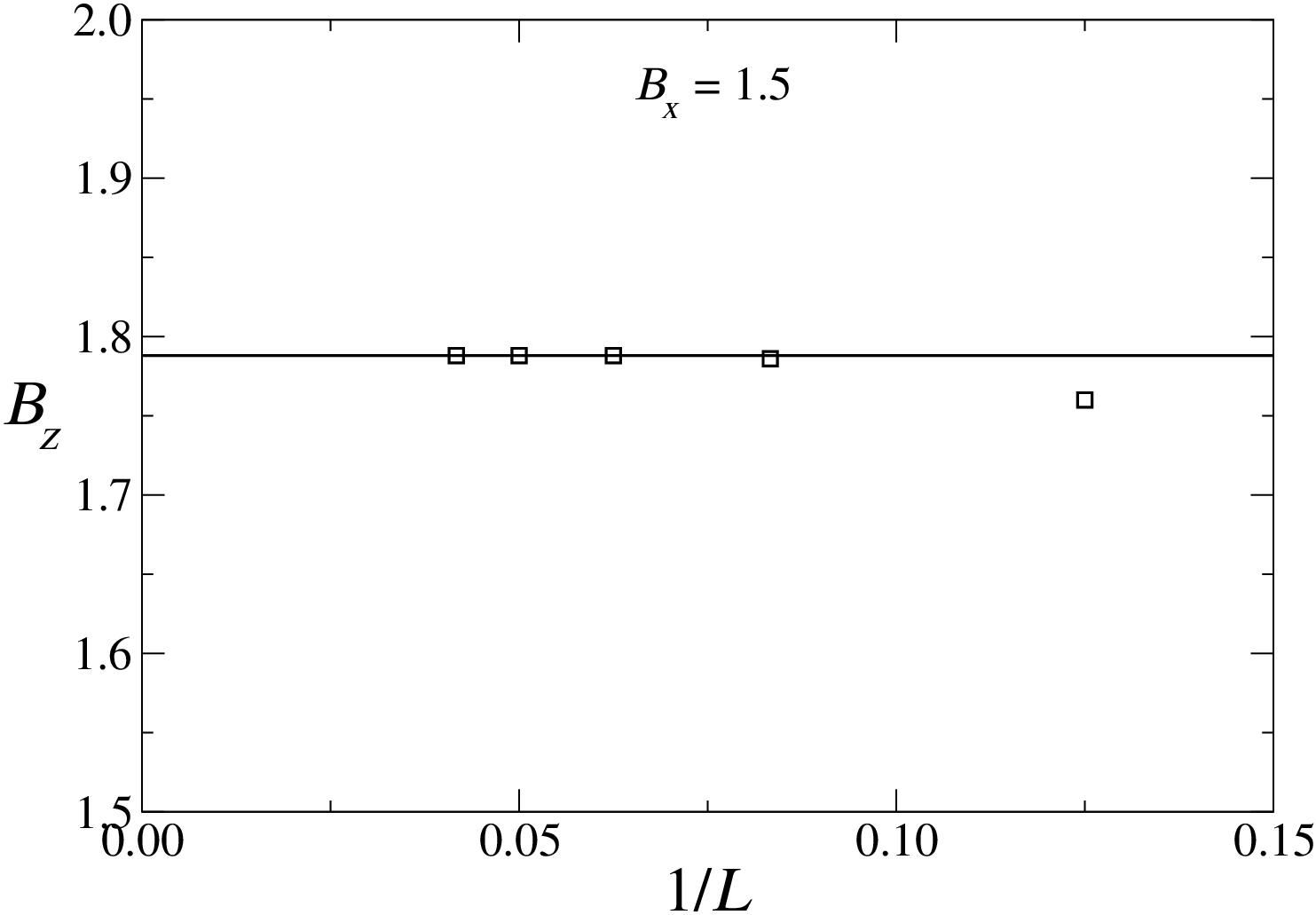}
\setlength{\abovecaptionskip}{0pt}
\setlength{\belowcaptionskip}{40pt}
\caption{(color online) Critical value $B_z$ at the transition line
between disordered and paramagnetic phases as a function of 
$1/L$ for $B_x=1.5$.  The extrapolated straight line to the
origin yields the thermodynamic value $B_z = 1.788$.  
\label{fig:figure5}
}
\end{figure}

\begin{figure}
\includegraphics[width=8.0cm, height= 6.0cm, angle=0]{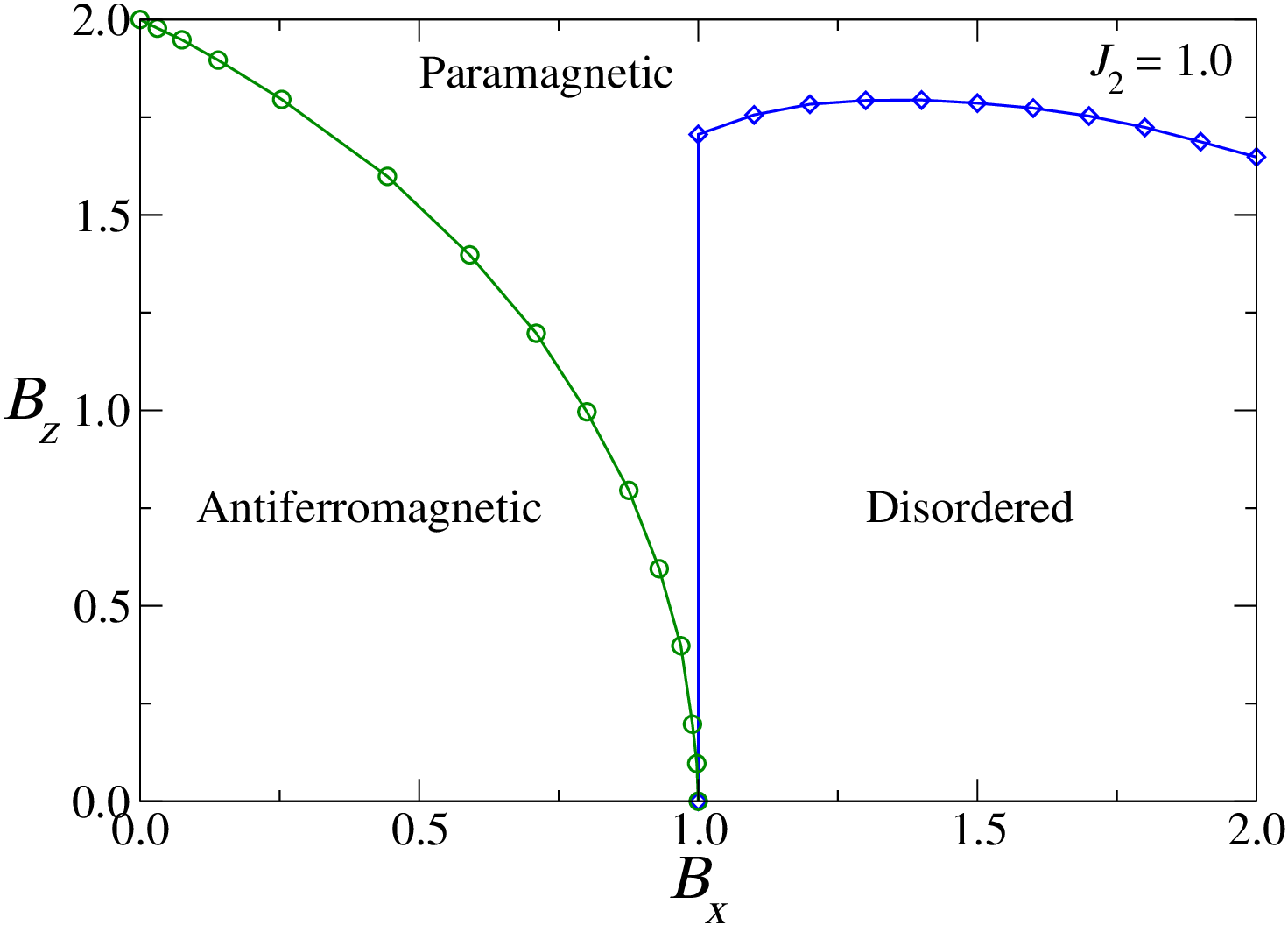}
\setlength{\abovecaptionskip}{20pt}
\setlength{\belowcaptionskip}{0pt}
\caption{(color online) Full phase diagram of the model in the $(B_x, B_z)$-plane using the results of the 
present work (diamonds)  and those of DMRG (open circles). The model shows three phase 
regions, antiferromagnetic, paramagnetic, and disordered.
The transition points  $(B_x,B_z) = (0, 2)$ and $(B_x,B_z) = (1, 0)$   are known exact results.
These phases are separated by second-order phase transitions.
\label{fig:figure6}
}
\end{figure}

\section{\label{sec:IV} Results} 

In our numerical calculations we used even lattice sizes from $L$ = 8 to 24. 
This choice of lattice sizes preserves the symmetry of the ground state when the system 
is in the antiferromagnetic phase. In addition, it avoids undesirable frustration effects due 
to the finite size of the system and the imposed periodic boundary conditions.

To obtain the phase diagram of the model, we first calculated the fidelity susceptibility as a function
of the transverse field $B_x$ for a fixed longitudinal field $B_z$. We represented this
susceptibility as $\chi_{B_z}$. In Fig.~{\ref{fig:figure1}} we show the behavior of  
$\chi_{B_z}$ for three lattice sizes $L=12$, $16$ and $24$, for the particular value 
of the longitudinal field $B_z = 0.5$. In all calculations involving the susceptibility, we have 
used $\delta$= 0.001. The maximum of the susceptibility for each lattice size is taken as the 
quantum transition point from antiferromagnetic to paramagnetic phases
for this particular value of longitudinal field. 

By carrying out such calculations for different values of longitudinal fields 
in the interval $(0,2)$, we obtained the phase diagram shown in Fig.~{\ref{fig:figure2}}. 
The results for $L = 12$  (open circles) , 16 (squares) and 24 (diamonds) are shown together 
with the critical boundary (dashed line) from~\cite{Ovc03} calculated using DMRG. As one 
can see,  by increasing the lattice sizes from $L = 12$ to $24$ the critical line from the fidelity 
method gradually approaches the DMRG results. The critical line for $L$ = 24 is almost 
indistinguishable from that of DMRG. This is the full phase diagram of the model, as reported 
in the literature~\cite{Ovc03, Sim11,Yus18}.

However, an analysis of the phase transitions for small longitudinal or transverses fields shows 
an inconsistency in the phase diagram of Fig.~\ref{fig:figure2}. 
For instance, for small transverse fields we expect an antiferromagnetic to paramagnetic 
phase transition boundary near $B_z = 2$. On the other hand, in the limit of small longitudinal 
fields, and based on the exact results for the transverse Ising model, we expect an 
antiferromagnetic to disordered transition near $B_x = 1$.

Another way to see this is that for low $B_x$ and high $B_z$, the spins should be pointing 
in the z-direction and for opposite case, namely low $B_z$ and high $B_x$, the spins should 
be pointing  in the x-direction. Thus these two configurations cannot be part of the same phase.
Therefore a phase boundary between the disordered-paramagnetic phase must be present in the
phase diagram, Fig.~\ref{fig:figure2}.

We will show below that by evaluating a second fidelity susceptibility
for a fixed transverse field ($\chi_{B_x}$) this missing phase boundary can be located.  
As in the case of Fig.~\ref{fig:figure2}, we first calculated the susceptibility $\chi_{B_x}$ 
as a function of $B_z$ for fixed values of $B_x$. Figure~\ref{fig:figure3} shows the results 
for $B_x = 1.5$ and $0.5$ (inset). The value $B_x$ = 0.5 lies within the  
antiferromagnetic phase, while $B_x$ = 1.5 is in the disordered phase. 

A point worth noticing in Fig.~\ref{fig:figure3} is the relatively high ratio between the 
two fidelity susceptibility maxima.
The susceptibility maximum across the transition from the antiferromagnetic phase to the
paramagnetic phase, shown in the inset, is about $35$ times larger than that of the maximum for
the disordered to paramagnetic phase of the main figure.  That is to be expected since the destruction
of the antiferromagnetic order produces a very small fidelity (overlap of the wavefunctions)
at the transition, hence a large susceptibility.  On the other hand, the overlap between
the disordered and the paramagnetic phases should be substantially larger, since no
particular spin ordering is being broken, causing the susceptibility peak to be much
less pronounced.  Perhaps that is the reason why the disorder to paramagnetic
transition has been overlooked in the treatments using other methods~\cite{Ovc03, Sim11,Yus18}. 

The phase diagram obtained using the maxima of $\chi_{B_x}$ for magnetic 
fields in the interval ($0 \le B_z, B_x \le 2$) for lattice sizes $L$ = 12, 16, and 24 is depicted 
in Fig.~\ref{fig:figure4}. 
For comparison, we have also included the DMRG results (dashed line).
The transition boundary between the antiferromagnetic and paramagnetic phases 
gets closer to the DMRG results as the chain size increases (although the convergence 
is slower for the susceptibility $\chi_{B_z}$).  Our fidelity results for the phase boundaries 
of different lattice sizes converge quite rapidly.  The boundary between 
the disordered and paramagnetic phases for $L$ = 16 and $L$ = 24 are already 
indistinguishable in the scale of the figure.
Figure~\ref{fig:figure5} illustrates the convergence of results for the critical field $B_z$ 
as we consider larger lattices.  In that figure, $B_x= 1.5$.  It  shows the values
of $B_z$  which maximize the susceptibility as a function of the inverse lattice size $1/L$.
As can be seen, the data converges rapidly to $B_z = 1.788$ at the thermodynamic limit. 
  Thus, the two fidelity susceptibilities  
 $\chi_{B_x}$ and $\chi_{B_z}$ complement each other in the determination 
of the phase boundaries. By combining the  results of the present work and those 
of DMRG, we arrived at the full phase diagram for the model,  depicted in Fig.~\ref{fig:figure6}.

\begin{figure}
\includegraphics[width=8.0cm,height= 6.0cm,angle=0]{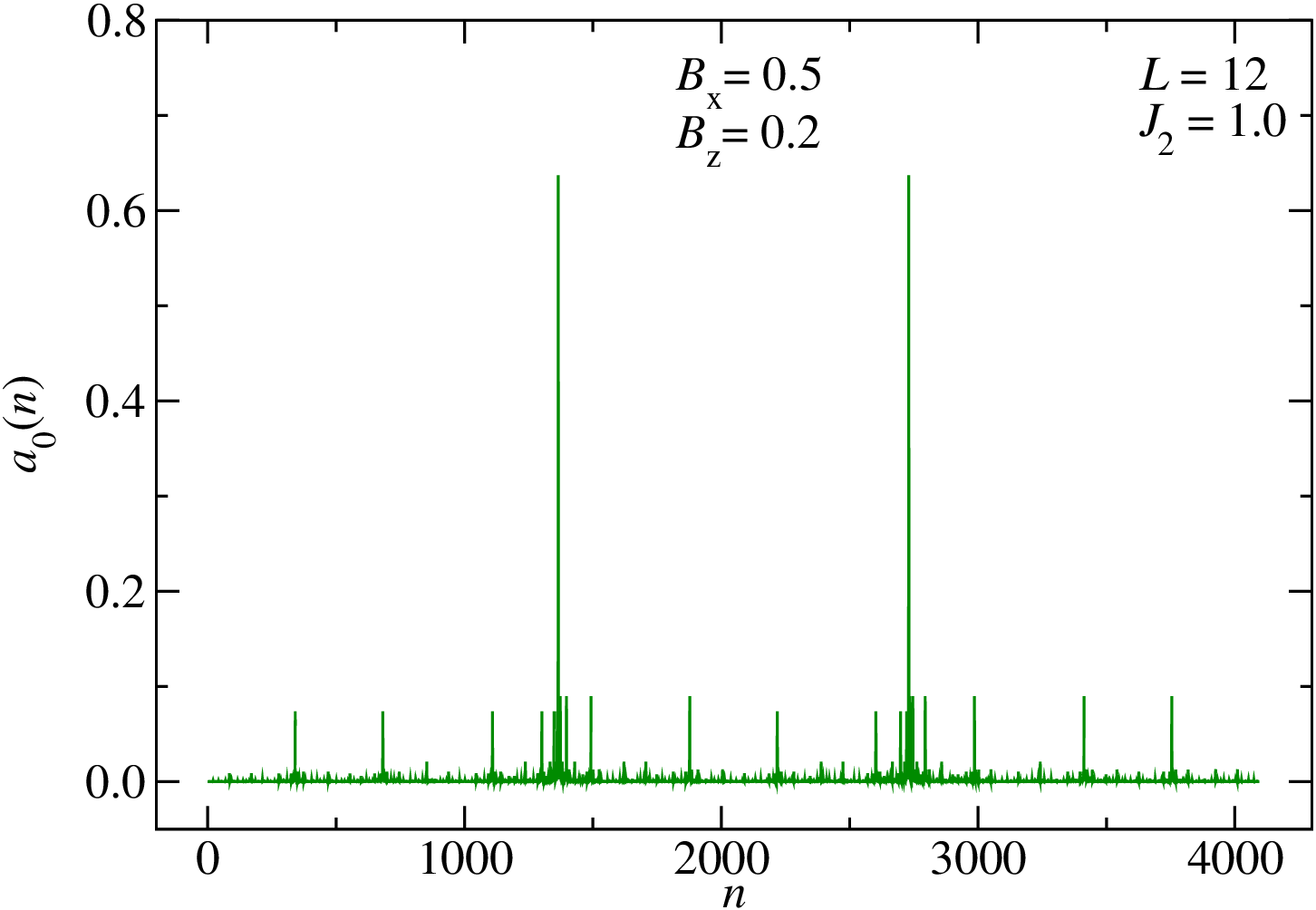}
\setlength{\abovecaptionskip}{0pt}
\setlength{\belowcaptionskip}{40pt}
\caption{(color online) Ground-state amplitudes {\em vs} the basis state 
index $n$ for $(B_x, B_z) = (0.5, 0.2)$, within the antiferromagnetic phase for $L = 12$.  
The two largest amplitudes correspond to an antiferromagnetic ordering. 
The smaller amplitudes are a signature of the transverse magnetic field. 
\label{fig:figure7}
}
\end{figure}

\begin{figure}
\includegraphics[width=8.0cm,height= 6.0cm,angle=0]{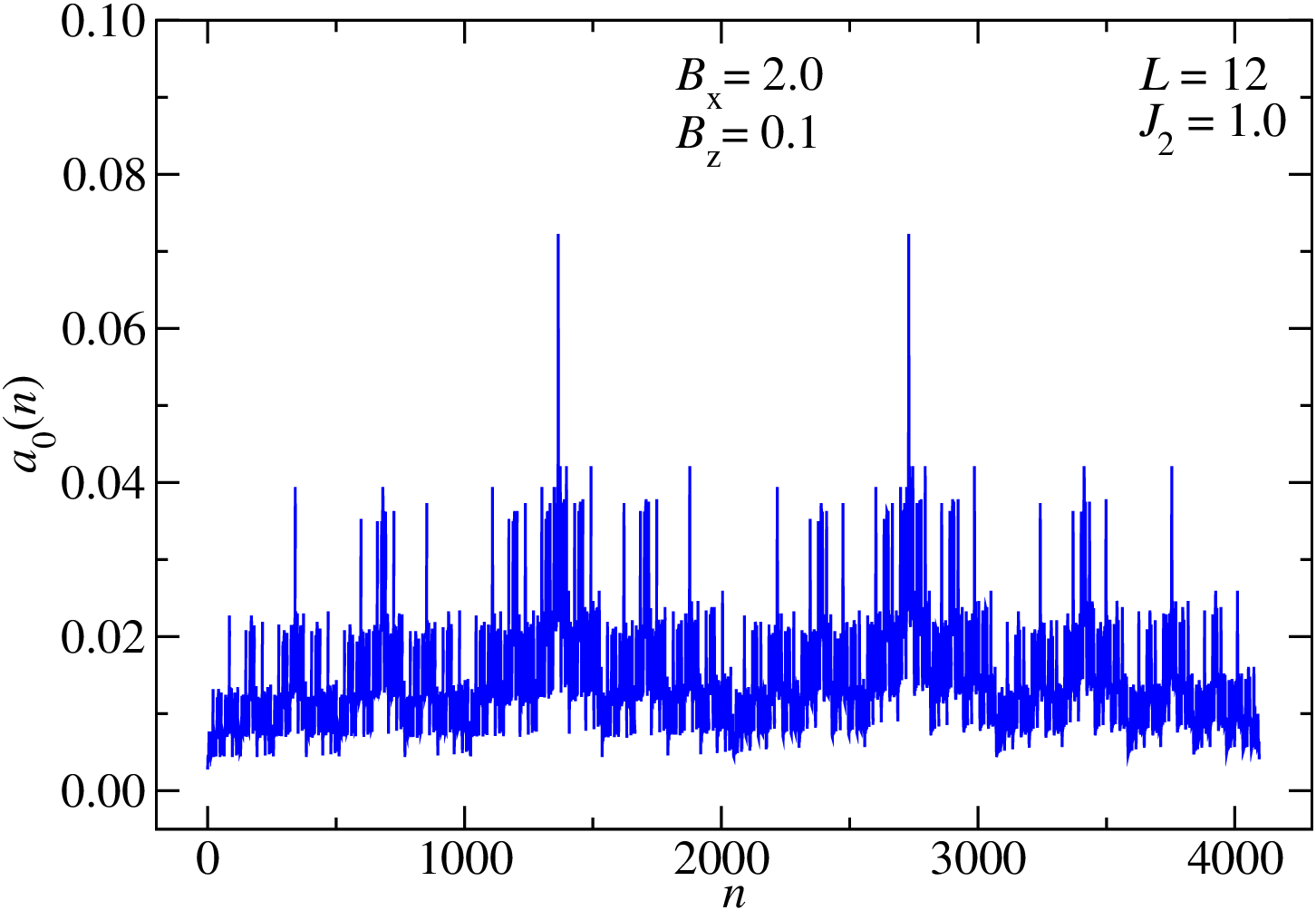}
\setlength{\abovecaptionskip}{-20pt}
\setlength{\belowcaptionskip}{60pt}
\caption{(color online) Ground-state amplitude {\em vs} basis state 
index $n$, for $(B_x, B_z) = (2.0, 0.1)$, in the disordered phase for $L = 12$.  
This spin configuration corresponds to a disordered phase.
The two largest amplitudes are for antiferromagnetic ordering. 
The smaller, yet comparable, amplitudes arise from quantum effects of the transverse 
magnetic field. 
\label{fig:figure8}
}
\end{figure}

\begin{figure}
\includegraphics[width=8.0cm,height= 6.0cm,angle=0]{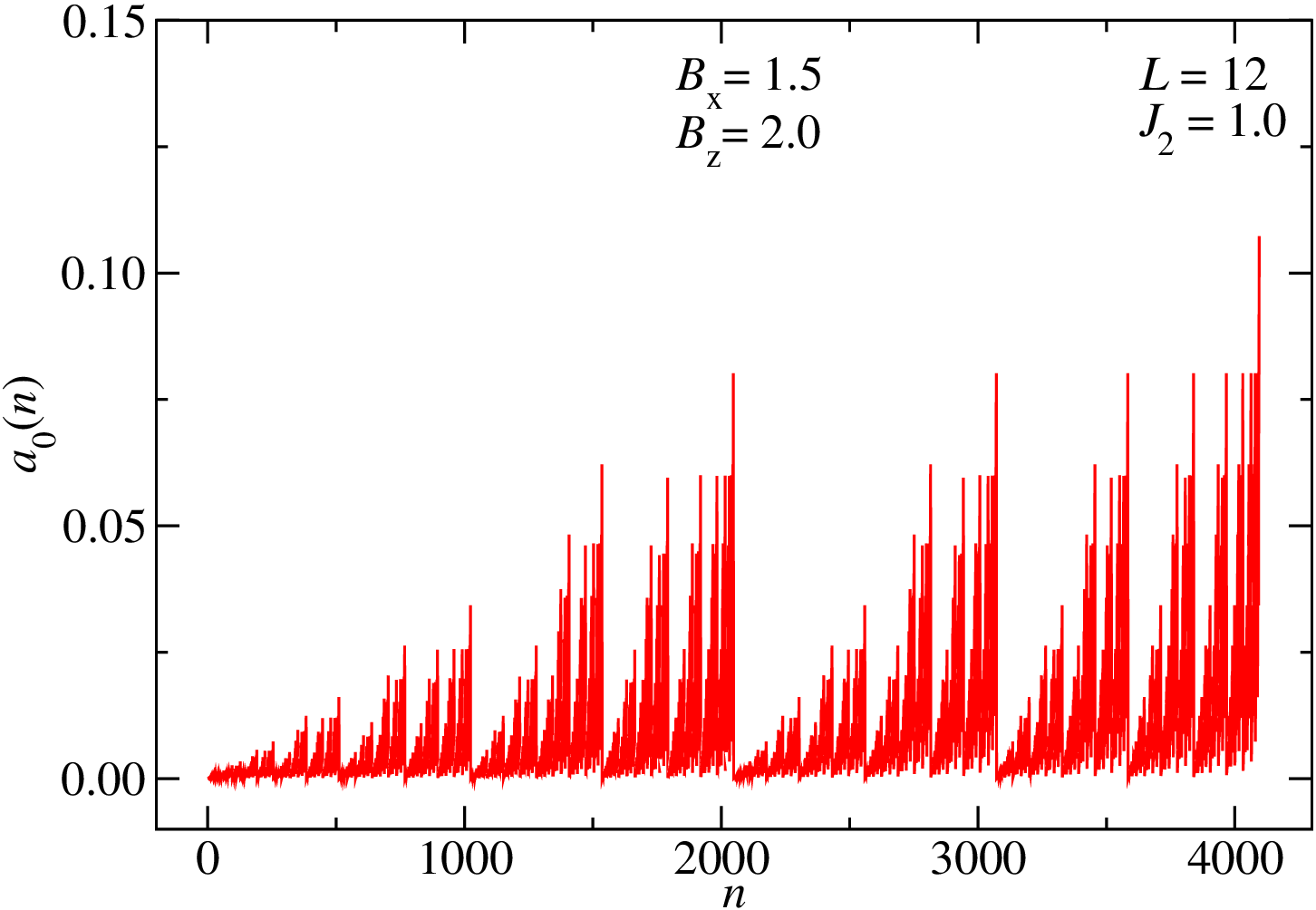}
\setlength{\abovecaptionskip}{-20pt}
\setlength{\belowcaptionskip}{0pt}
\caption{(color online) Ground-state amplitudes {\em vs} basis state 
index $n$ for $(B_x, B_z) = (1.5, 2.0)$, within the paramagnetic phase 
for $L = 12$.  
The largest amplitude corresponds to the ferromagnetic configuration. 
\label{fig:figure9}
}
\end{figure}

The spin configuration of each phase can be found by plotting the ground-state eigenvector 
amplitudes as a function of the ground-state index $n$. As a working example we shall 
use $L$ = 12. For the point $(B_x,B_z)$= (0.5, 0.2) inside the antiferromagnetic phase, we 
obtained the plot depicted in Fig.~\ref{fig:figure7}. The two largest amplitudes are at  
$n$=1365 and $n$=2730, corresponding to a ground-state in the binary representation 
$|$010101010101$>$ and $|$101010101010$>$, respectively. The much smaller amplitudes 
are transverse field effects. 

Moving to the disordered phase, we consider now the point $(B_x,B_z) = (2.0, 0.1)$, where
the ground-state amplitudes are shown in Fig.~\ref{fig:figure8}. 
Although the antiferromagnetic component is still present (due to the Ising interactions)
as the larger component of the amplitudes,  many other components with comparable 
amplitudes are also present.

Finally, we considered $(B_x,B_z) = (1.5, 2.0)$, inside the paramagnetic phase.  We obtained 
the graph shown in Fig.~\ref{fig:figure9}. The largest amplitude at $n= 4096$ corresponds 
to the ferromagnetic configuration with all spins pointing in the direction of the field. 
The second largest amplitudes also correspond to a ferromagnetic configuration with all spins 
but one aligned with the field. The third largest amplitudes still correspond to a ferromagnetic 
configuration with all spins but two aligned with the field.  Similar ferromagnetic configurations 
are found for the smaller amplitudes.

\vskip 1 truecm

\section{\label{sec:V} Summary and Conclusions}

The ground-state properties of the transverse Ising model in the presence of a 
longitudinal field was analyzed using the quantum fidelity method. The phase diagram 
in the $(B_x,B_z)$-plane shows three phases in contrast with previously reported results 
from the literature, which show only two phases. The phases are antiferromagnetic, 
paramagnetic, and disordered. These phases are separated by second-order phase 
transitions. We have also analyzed the spin configuration of the ground-state of each 
corresponding phase. The spins configuration on each phase clearly show 
distinct characteristics.

\vskip 24pt
\centerline{\bf ACKNOWLEDGEMENTS}
\vskip 24 pt
We thank C. Warner for critical reading of the manuscript.
O.F.A.B. acknowledges support from the Murdoch
College of Science Research Program and a grant from the 
Research Corporation through the Cottrell College Science 
Award No.~CC5737.
We also thank FAPERJ, CNPq and PROPPI (UFF)  for financial support.

\end{document}